# Computational coherent averaging for free-running dual-comb spectroscopy

LUKASZ A. STERCZEWSKI,[1,2,3,†] JONAS WESTBERG,[2,†] AND GERARD WYSOCKI[2,*]

[1]*Department of Electrical Engineering, Princeton University, Princeton, New Jersey 08544, USA*
[2]*Faculty of Electronics, Wroclaw University of Science and Technology, Wroclaw 50370, Poland*
[3]*Kosciuszko Foundation Visiting Scholar residing at Princeton University*
[†]*These authors contributed equally to this work.*
*\*gwysocki@princeton.edu*

**Abstract:** Dual-comb spectroscopy is a rapidly developing technique that enables moving parts-free, simultaneously broadband and high-resolution measurements with microseconds of acquisition time. However, for high sensitivity measurements and extended duration of operation, a coherent averaging procedure is essential. To date, most coherent averaging schemes require additional electro-optical components, which increase system complexity and cost. Instead, we propose an all-computational solution that is compatible with real-time architectures and allows for coherent averaging of spectra generated by free-running systems. The efficacy of the computational correction algorithm is demonstrated using spectra acquired with a THz quantum cascade laser-based dual-comb spectrometer.

2018-05-21

**1. Introduction**

Dual-comb spectroscopy (DCS) [1–3] derives its strength from the massively parallel heterodyne down-conversion procedure that enables direct mapping of the information encoded in the optical domain to the radio-frequency (rf) domain, where high-speed analog-to-digital converters (ADCs) can be used to acquire the signals. This requires spatial overlap of the beams from two matched frequency comb sources onto a fast photodetector, causing the optical beating frequencies of the modes to be uniformly spread over the detector bandwidth. Unlike most broadband techniques, this gives nearly instantaneous access to the available optical bandwidth, with applications in environmental monitoring, industrial process control and characterization of transient chemical dynamics, where the combination of spectral coverage and time resolution can be effectively leveraged. However, a stringent requirement is placed on the mutual coherence of the two comb sources, and any significant drift or fluctuation in phase will degrade the system performance over longer time-scales. This effect can be mitigated if the phase noise can be actively suppressed through a high-bandwidth hardware feedback loop, or measured continuously to phase-shift the acquired signals and adaptively adjust the ADC sampling rate. If implemented successfully, the time-dependent distribution of the spectral power will be aligned in the frequency domain enabling averaging without dispersing the power over a large rf bandwidth, so called coherent averaging, also known in radar systems as coherent integration. It is important to note that for some applications a straightforward alignment of multiple short-time DCS magnitude spectra [4–6] or time-domain interferogram bursts [7] may be sufficient. Nevertheless, there are some inherent disadvantages of this approach: the linewidths of the beat notes are limited by the acquisition time of a single frame and the lack of repetition rate correction does not account for the cumulative broadening of the beat notes at the edges of the spectrum. In addition, only beat notes identifiable within a single frame can be used, which precludes sporadically appearing weaker beat notes, thus limiting the operational bandwidth.

The conventional approach for coherent averaging in DCS systems incorporates additional electro-optical hardware, such as additional external cw reference lasers and photodetectors [8,9]. While such an approach is feasible in the near-infrared where the developments in telecommunication components can be leveraged, at longer wavelengths (mid- to far-infrared), low manufacturing quantities decrease the availability and increase the prices. For this reason, we propose a purely computational coherent averaging (CoCoA) algorithm that is compatible with real-time architectures and acts directly on the digitized photodetector signal. This allows for correction of relative phase drifts and fluctuations even when the rf beat notes are masked by excessive phase noise. Although the procedure can be applied to any free-running DCS system, e.g. such based on electro-optic modulators, optically pumped microresonators, or fiber lasers, the main motivation for this work is linked to the recent developments in chip-scale frequency combs based on electrically-pumped semiconductor sources such as quantum cascade lasers (QCLs) [4,10–14] or interband cascade lasers (ICLs) [15,16]. These sources have begun to reach a state of development where they can be considered as the basis for miniaturized, battery-operated, portable spectrometers with small footprints [17], but unfortunately this miniaturization comes at a price. Free-running semiconductor combs exhibit fluctuations in repetition rate ($f_{rep}$) on the order of hundreds of hertz, which compared to typical repetition rates of several gigahertz, corresponds to a relative stability of ~$10^{-7}$. This is several orders of magnitude worse than that seen in fiber-based combs, which routinely reach relative stabilities of <$10^{-9}$. In addition to the repetition rate, the comb offset frequency ($f_0$), commonly referred to as carrier-envelope offset (CEO) frequency, fluctuates as well. These two sources of phase noise are schematically shown in Fig. 1.

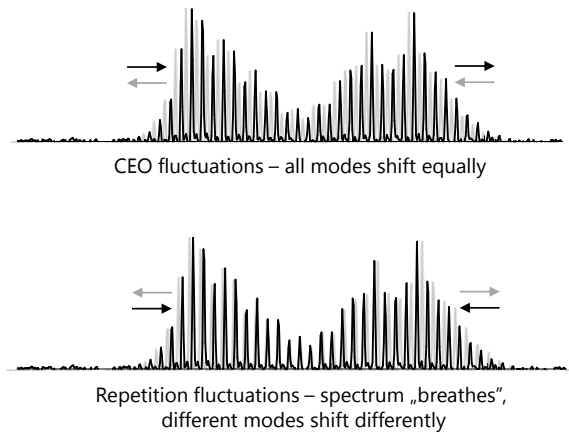

CEO fluctuations – all modes shift equally

Repetition fluctuations – spectrum „breathes",
different modes shift differently

Fig. 1. Frequency instabilities in frequency combs. a) CEO fluctuations b) Repetition rate fluctuations.

Even though such combs show high mode-to-mode coherence, they are corrupted by considerable amounts of phase noise. When such free-running combs are spatially overlapped on a fast photodetector, rf beat notes with linewidths approaching MHz-level are observed. In the worst case, instead of evenly spaced down-converted narrow rf comb lines, the rf power will be dispersed over the entire bandwidth, as shown in Fig. 2a. Clearly, a scenario like this is incompatible with DCS, but the apparent loss of information can be remedied via proper phase correction.

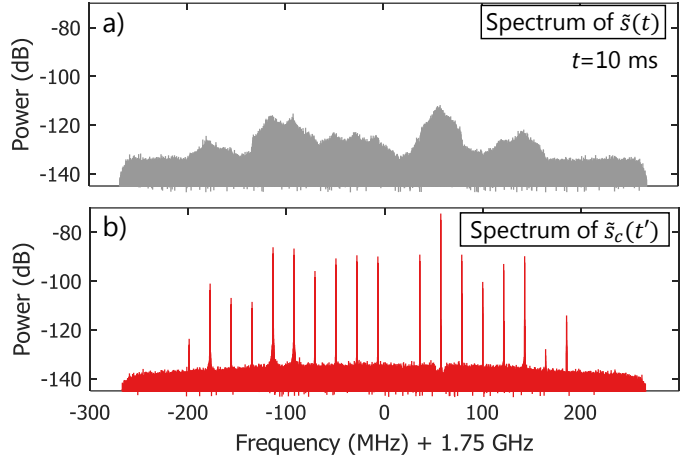

Fig. 2. Demonstration of the efficacy of the coherent averaging algorithm. a) Raw dual-comb rf spectrum acquired using two free-running terahertz quantum cascade lasers. b) rf spectrum after timing and phase correction.

A well-known fact about the Fourier Transform states that it is not suitable for analysis of non-stationary signals, i.e. such that change their statistics over the acquisition time [18]. Beat notes whose frequencies fluctuate randomly (often much more than the spacing between the lines) are undoubtedly examples of such non-stationarity (see Fig. 2a). To make matters worse, each rf beat note has a unique frequency deviation and hence unique phase noise characteristics. However, by taking advantage of the underlying comb nature of the laser source, the beat notes can be effectively recovered from the noise and used in practical applications. Recent works on digital correction algorithms [19,20] clearly show this capability. In 2016, Burghoff et al. presented the first demonstration of a purely computational correction of DCS signals [19]. Their approach relied on the Extended Kalman Filter (EKF), which reconstructed the amplitudes, phases, and frequencies of all RF beat notes in the DCS spectrum. Although effective, the computational complexity of the EKF caused predominantly by the inversion of a large matrix, limits its application to systems with less than a few dozens of modes.

Another approach proposed a year later by Hébert et al. involves the use of the cross-ambiguity function developed initially for radar applications [20]. Conceptually, it is similar to the well-known cross-correlation technique capable of retrieving the time delay between consecutive burst of the interferogram and hence retrieve the repetition rate, while in addition the frequency offset between two similar waveforms can be obtained. An impressive correction of approximately three thousand comb teeth initially dispersed in the RF domain due to $f_0$ and $f_{rep}$ fluctuations was demonstrated. Unfortunately, the algorithm requires the generation of a two-dimensional map with multiple points for each burst of the interferogram, followed by a search for the global maximum at each step. This procedure is computationally demanding and not well-suited for real-time correction. Also, systems where no identifiable zero path difference bursts appear in the beating signal, such as QCL- and ICL-based DCS systems [5,15], are not compatible with this approach.

Our computational coherent averaging (CoCoA) approach mitigates most of those limitations, while being dramatically less computationally complex. Firstly, it does not require the presence of any interferogram bursts or identifiable frames. Instead, it retrieves the instantaneous repetition rate through a simple single-step non-linear operation: the sum of the squared real and complex components, followed by coherent demodulation of any harmonic using an IQ demodulator, which is easily realizable in programmable hardware. This signal acts as an adaptive clock to keep the spacing between the rf comb lines equidistant. Once this is ensured, we employ a simple recursive formula frequency tracker, which again is easy to implement in hardware and corrects for the global offset phase noise. Consequently, we correct for both sources of instabilities in the down-converted rf comb. Most importantly, the limitation of a single frame resolution does not pertain here. The linewidth of the corrected rf comb lines is mainly limited by the acquisition time, provided the sources show sufficient stability at a $1/\Delta f_{rep}$ timescale, and that the extraction of the correction signal is not corrupted by the noise of the photodetector. Another feature of the CoCoA algorithm is that its output is a phase-corrected time-domain signal, which allows the use of ordinary rf filters for analyzing data in swept spectroscopy mode [21]. The algorithm shows almost no degradation of correction efficacy for higher frequency rf beat notes, ascribable to the use of high-order harmonics of the repetition rate, and its computational complexity does not depend on the number of comb lines in the spectrum. Finally, even the weakest beat notes can be used for spectroscopy because the entire duration of the beating signal can be harnessed for the calculation of the Fourier Transform. It should be noted that the CoCoA algorithm works directly on the acquired interferograms and does not correct the phases of the optical modes interacting with the sample. It merely assures mutual coherence by manipulating the timing and phase of the acquired photodetector signal.

## 2. Short-time Fourier Transform analysis of radio-frequency spectra

Before we proceed to an actual description of the CoCoA algorithm, let us analyze the effect of acquisition time on the shape of the acquired DCS spectrum as illustrated by the data shown in Fig. 3. In this experiment we optically beat two cryogenically-cooled terahertz QCL phase-noisy combs on a 7.5 GHz bandwidth superconducting hot electron bolometer (Scontel, Russia), and acquire rf data by IQ-demodulating at 1.75 GHz using a real-time spectrum analyzer (R&S FSW-43). This example is selected to illustrate the strengths of the CoCoA algorithm due to the strong presence of phase noise. Also, the spectrum contains a relatively low number of rf beat notes, which make the improvements clearly visible. However, the algorithm works equally well for a larger set of beat notes, of which examples are given in section 5. The top panel of Fig. 3 shows the frequency spectrum at 1 μs, wherein a dozen of strong beat notes is easily distinguishable from the large-variance noise floor. The 3 dB linewidth is close to the Fourier limit defined by the acquisition time (~1 MHz), which confirms that the two combs are stable at a $1/\Delta f_{rep}$ timescale, unfortunately accompanied by the typical uncertainty of the peak amplitude on the order of tens of percent [15,22]. In other words, a transmission DCS measurement using such a short acquisition time would be associated with large error bars. Naively, one would expect that by increasing the acquisition time ten times (10 μs), a significant increase in the signal-to-noise ratio would be observed. Unfortunately, the opposite is true, which stems from a broadening of the beat notes that take highly asymmetric shapes. A further increase by an additional factor of ten (100 μs) causes the beat notes to disperse over tens of megahertz. Naturally, the retrieval of peak amplitudes for spectroscopic assessments with such non-uniform shapes is not a trivial task.

One way to analyze non-stationary signals with the Fourier Transform is to calculate it over shorter time scales when the considered signal is locally stationary, commonly referred to as Short-time Fourier Transform (STFT). A plot of the squared magnitude of the STFT as a function time is known as the spectrogram, which is shown in Fig. 3b. The spectrogram is generated for 1 ms of acquisition time with a temporal resolution of 1 μs. The oscillatory fluctuations of the beat notes are caused by the noisy environment of the vibrating cryostat, as well as the presence of optical feedback in the system corrupting stable comb operation. It is important to note that in general the properties of the fluctuations are much more complex without a periodic character. This example clearly demonstrates that ordinary averaging fails to be effective when applied to DCS signals affected by phase noise. Therefore, it is imperative to apply the appropriate corrections in order to suppress the noise and increase the signal-to-noise for low-uncertainty amplitude estimation.

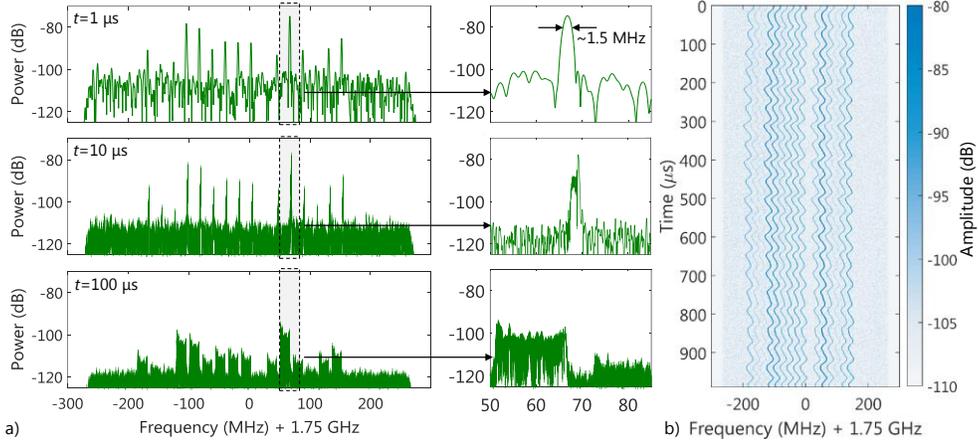

Fig. 3. Effect of acquisition time on the shape of the DCS spectrum. a) RF spectra at 1 μs, 10 μs, and 100 μs together with a zoom on the strongest beat note. Only the shortest time acquisition yields a nearly pure, acquisition time limited tone. Longer time acquisitions reveal the multi-peak complex shape of the beat notes. (b) Spectrogram of a 1 ms acquisition of two QCLs suffering from cryostat vibrations and optical feedback. $1/\Delta f_{\text{rep}}$ timescale.

## 3. Phase and timing correction for coherent averaging

At the core of the CoCoA algorithm lies the observation that once the mode-number-dependent phase noise contribution of the fluctuating combs' differential repetition rates has been removed, we need only to correct for common relative phase noise. Therefore, it is sufficient to resample the signal to compensate for a variable duration of the consecutive beating interferograms, followed by a global tracking and correction of the difference in offset frequency. A block diagram of the proposed solution is shown in Fig. 4.

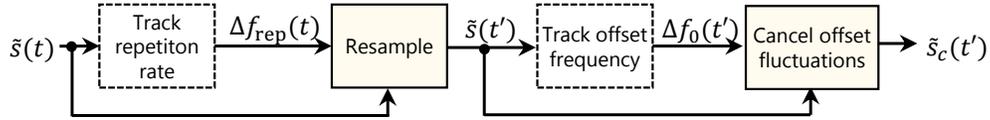

Fig. 4. Block diagram of the proposed timing and phase correction.

The algorithm consists of two estimation and two signal manipulation blocks, included in dashed, and solid boxes, respectively. We will provide a general description of each signal manipulation block in the following subsections and additional details related to the signal processing performed in each block are provided in the appendix.

### 3.1. Tracking of the instantaneous repetition rate

For the sake of simplicity, consider a complex dual-comb signal $\tilde{s}(t)$ consisting of $N$ unitary amplitude components with an rf offset frequency $\Delta f_0$ and an rf repetition rate $\Delta f_{\text{rep}}$

$$\tilde{s}(t) = \exp(i2\pi\Delta f_0 t) \sum_{n=1}^{N} \exp(i2\pi n\Delta f_{\text{rep}} t). \quad (1)$$

As shown in the appendix, self-mixing followed by combining the squared real and imaginary parts enables an extraction of the harmonics of the repetition rate without any offset frequency components,

$$(\text{Re}\{\tilde{s}(t)\})^2 + (\text{Im}\{\tilde{s}(t)\})^2 = \underbrace{N}_{\text{DC}} + 2\underbrace{\sum_{m=1}^{N}\sum_{l=1}^{m-1}\cos(2\pi(m-l)\Delta f_{\text{rep}}t)}_{\text{DFG}}. \qquad (2)$$

This formula, referred here to as digital difference frequency generation (DDFG), produces a signal that is the result of the beating between all possible rf beat notes, i.e. a superposition of beating products between neighboring beat notes is obtained (every other, third etc.). Obviously, higher order harmonics of the fluctuating repetition rate will be more dispersed in frequency than the lower ones due the cumulative broadening effect, which increases linearly with harmonic number.

Once the DDFG signal is obtained, the instantaneous frequency $k\Delta f_{\text{rep}}(t)$ of the highest-order rf repetition rate harmonic $k$ can be tracked with a high signal-to-noise ratio (in practice ~20 dB). The tracking procedure can be realized through computational IQ demodulation at the expected value of the repetition rate harmonic within a fixed bandwidth, by employing filtering and the Hilbert transform, or by implementing more sophisticated harmonic frequency trackers [23], as described in the offset frequency paragraph. Frequency trackers are capable of handling overlapping beat notes, which is often encountered in combs with more than a hundred spectral modes and small repetition rate difference.

The DDFG operation also serves as an important diagnostic tool. A DCS signal that shows no repetition rate harmonics is non-correctable using the procedure described here because it does not possess comb characteristics. This useful feature will be discussed in Section 5.

### 3.2. Correction of the variable repetition rate

Having retrieved the $k$-th harmonic of the instantaneous rf comb repetition rate $k\Delta f_{\text{rep}}(t)$, the time-domain signal needs to be resampled onto a non-uniform grid to cancel the effects of the repetition rate fluctuations. The time axis used for resampling has a cumulative character, and is given in the continuous case by

$$t'(t) = \int_0^t \frac{k\langle\Delta f_{\text{rep}}\rangle}{k\Delta f_{\text{rep}}(\tau)}\,d\tau, \qquad (3)$$

where $\langle\Delta f_{\text{rep}}\rangle$ denotes the expected value of the repetition rate. The resampling operation aims to ensure an equal duration of the beating events in the interferogram. In this example, we are using the most trivial linear interpolation between two sampled points. Note that in the case of semiconductor combs equipped with a microwave bias tee, one can electrically mix the intermode beat notes and feed the difference frequency to the data acquisition reference clock, thereby avoiding this procedure.

### 3.3. Tracking of the instantaneous offset frequency

After the repetition rate correction that ensures constant spacings in the down-converted rf comb, the global offset frequency fluctuations need to be extracted. In principle, all beat notes in the rf spectrum can be tracked using a modified Multiple Frequency Tracker [24] or the Kalman Filter [19]. However, after the repetition rate correction, the complexity of this task can be significantly reduced to tracking of only one beat note that can be isolated using a digital band-pass filter. We have identified a few schemes for this purpose with different complexity, yet with nearly the same tracking performance. The simplest utilizes a recursion approach for a real discretely-sampled signal known as the Fast Frequency Tracker, proposed by R. Aarts [25], who showed that the formula

$$\hat{r}_k = \hat{r}_{k-1} + x_{k-1}\gamma[x_k + x_{k-2} - 2x_{k-1}\hat{r}_{k-1}] \qquad (4)$$

can be used to track the instantaneous frequency of a slowly-varying non-stationary sinusoidal signal sampled in $x_k$ with a forgetting factor $\gamma$. Of course, indices $k-1$ and $k-2$ denote values from previous time instances. Since $\hat{r}_k$ is an auxiliary signal that contains the tracked frequency in the argument of the cosine

$$r_k = \cos(2\pi \Delta f_0(k) T_s), \qquad (5)$$

where $T_s = 1/f_s$ is the sampling period, to estimate the instantaneous frequency $\Delta f_0$ at $k$-th iteration we need to calculate the arccosine of $\hat{r}_k$ scaled by the sampling frequency and $1/2\pi$

$$\Delta \hat{f}_0(k) = \cos^{-1}(\hat{r}_k) \cdot \frac{f_s}{2\pi}. \qquad (6)$$

The tracking formula $\hat{r}_k$ uses only multiplications and additions, therefore it is compatible with a real-time platform. Also, the arccosine function can be tabulated to facilitate real-time offset frequency tracking.

One of the drawbacks of the Fast Frequency Tracker is a degradation of its performance when $x_k$ is composed of numerous sine waves, which occurs when the drift of the offset frequency exceeds the spacing between the rf beat notes. We have found two solutions to effectively solve this problem. The first employs the aforementioned Multiple Frequency Tracker [24], which is proven to be robust while having a very low tracking delay, at the expense of computational complexity due to its use of matrix inversions. The second is inspired by carrier recovery systems in wireless networks, which comprise a coarse and fine carrier frequency tracker, where the fine tracker is simply the previously discussed Fast Frequency Tracker. The coarse estimation keeps the band-pass filter for $x_k$ at the pertinent frequency and is realized through the Short-time Fourier Transform and cross-correlation detecting the shift in the position of the magnitude spectrum by the same token as in [5]. In short, we reduce the problem of drifts exceeding the spacing between the rf lines to tracking residual fluctuations around almost stationary beat notes. The use of the STFT is favored due to its real-time compatibility with modern signal processing platforms. A review of different frequency trackers together with their derivation and implementation details is given in [26].

### 3.4. Correction of the variable offset frequency

The final step in the phase correction is a complex multiplication to counteract the offset frequency fluctuations. The following formula multiplies the adaptively-resampled signal by the offset phase fluctuations in counterphase,

$$\tilde{s}_c(t') = \exp\left(-j2\pi \int_0^t \Delta f_0(\tau) - \langle \Delta f_0 \rangle d\tau \right) \cdot \tilde{s}(t'). \qquad (7)$$

The $\langle \Delta f_0 \rangle$ term ensures that the corrected spectrum remains at a constant noise-free position, which would otherwise be unnecessarily shifted by the correction procedure.

### 3.5. Flowchart of the CoCoA algorithm

To illustrate how the THz DCS spectrum is corrected, a flowchart of the signals after each step are given in Fig. 5. Fig. 5a shows the uncorrected IQ-demodulated photodetector signal after applying the FFT. As can be seen, the beat notes are not resolvable for the acquisition time used, in this case 1 ms. The results of the DDFG procedure, described in Section 3.1, and the appendix, is shown in Fig. 5b. The 8[th] harmonic (white arrow) is chosen for demodulation and the retrieved instantaneous differential repetition rate is displayed in Fig. 5c. The repetition rate correction is applied according to Eq. (3) and the resulting spectrum is shown in Fig. 5e. If the resampling is effective, the resampled signal should demonstrate an increase in signal-to-noise for the harmonics after the DDFG procedure. This is shown in Fig. 5d, which compared to Fig. 5b has an average amplitude increase of ~20 dB.

The frequency tracking part of the algorithm is applied on the adaptively-sampled signal of Fig. 5e, after which the instantaneous difference frequency offset is retrieved. The complex multiplication of Eq. (7) finally gives the corrected rf spectrum shown in Fig. 5g. Fig. 5h shows the effect of applying only the frequency offset correction, which clearly corrects for the majority of the phase noise.

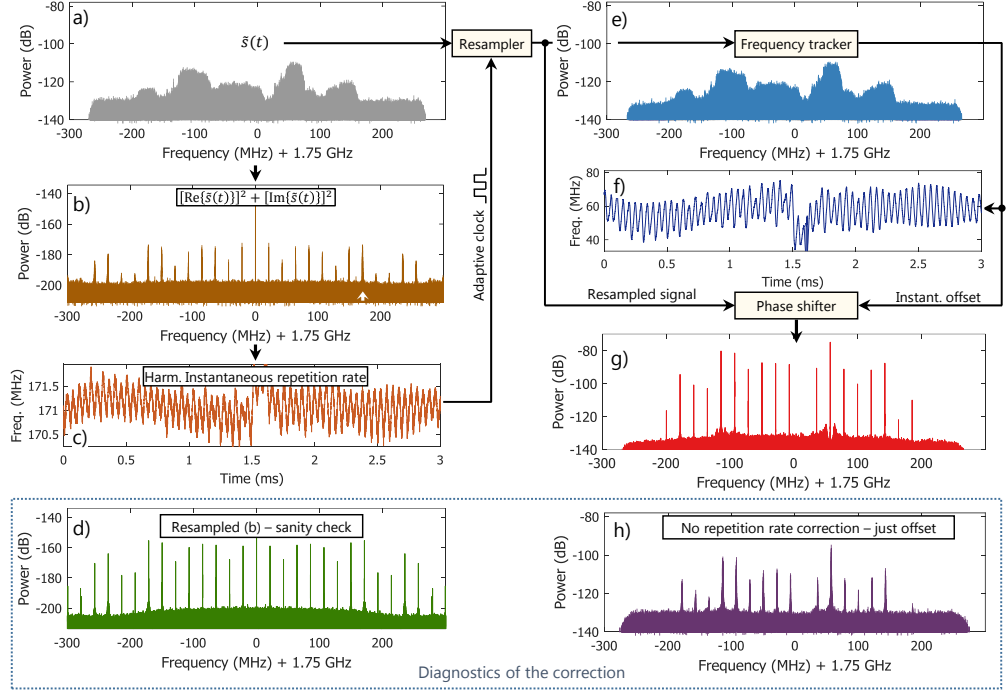

Fig. 5. Flowchart of signals in the CoCoA algorithm. (a) the IQ-demodulated DCS rf signal. (b) DDFG self-mixing spectrum. (c) the retrieved instantaneous differential repetition rate. (d) the resampled DDFG self-mixing spectrum. (e) the IQ-demodulated DCS rf signal after adaptive resampling. (f) the instantaneous difference offset frequency obtained from the fast frequency tracker. (g) the IQ-demodulated rf spectrum after correction. (h) the effect of solely correcting for difference frequency offset.

## 4. CoCoA as a diagnostic tool for incoherence detection

As mentioned earlier, the CoCoA algorithm can be used as a diagnostic tool to indicate the mutual coherence of the two sources contributing to the interferograms. To illustrate this property, two interband cascade lasers [16] centered at ~3.5 µm were used in the DCS configuration, and operated in the comb-regime or high phase-noise regime by altering the bias current. Fig. 6a shows the rf spectrum acquired when operating one of the lasers in the high-phase noise regime, indicated by the broad linewidth of the intermode beat note shown in Fig. 6d. The other laser was operated as a comb with a sub-kilohertz linewidth visible in Fig. 6c. Similar to Fig. 5a, the rf spectrum of Fig. 6a shows no resolvable beat notes, but as discussed earlier this could be the result of a long acquisition time. However, unlike the previously presented data, the self-mixing spectrum shown in Fig. 6b lacks the characteristic $\Delta f_{rep}$-harmonics seen whenever sufficient mutual coherence is present. Consequently, the CoCoA algorithm does not provide any improvement in the spectrum as the requirement on the input signal is not satisfied. In fact, the beat notes will become even more dispersed as there is no correlation between the frequency components of the signal. In contrast, when operating both lasers in the comb-regime, the self-mixing spectrum of 7b is obtained. The spectrum exhibits multiple $\Delta f_{rep}$-harmonics, any of which can be IQ-demodulated for the adaptive sampling step described in Section 3.2. The result of the CoCoA algorithm is apparent when comparing Fig. 7a and c. Again, the algorithm provides linewidths that are close to the Fourier limit imposed by the acquisition time (shown in Fig. 7e and f). The diagnostics provided by the self-mixing spectrum can be a useful tool when access to the intermode beat note, either through optical detection or rf extraction, is not available.

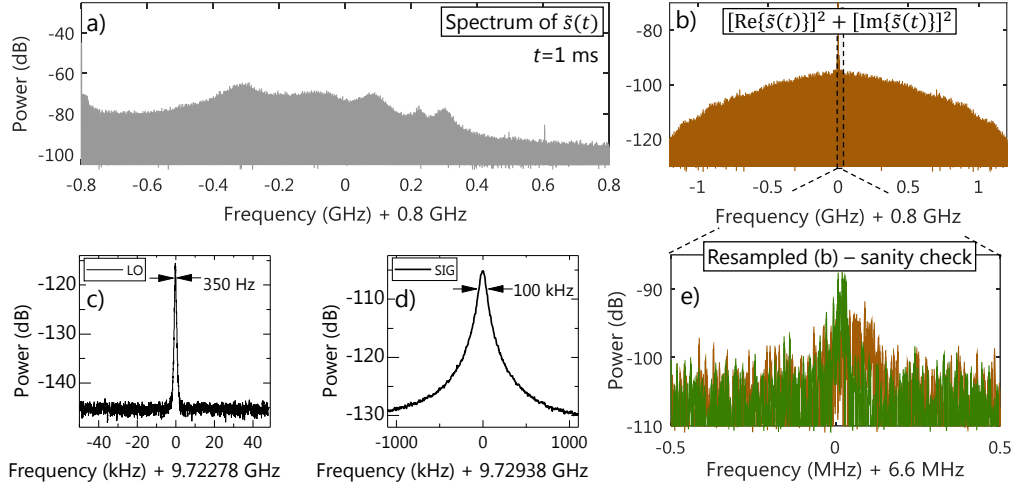

Fig. 6. (a) Dual-comb spectrum of two ICL combs acquired over 1 ms, where one is operated in the high phase-noise regime. As a result, no resolvable rf beat notes are observed. (b) the DDFG self-mixing spectrum with no indication of $\Delta f_{\text{rep}}$-harmonics, which is a signature of non-comb operation of at least one of the devices. (c) the intermode beat note of the device operating in comb-mode. (d) the intermode beat note of the device operating in high phase-noise regime. (e) DDFG self-mixing spectrum after resampling.

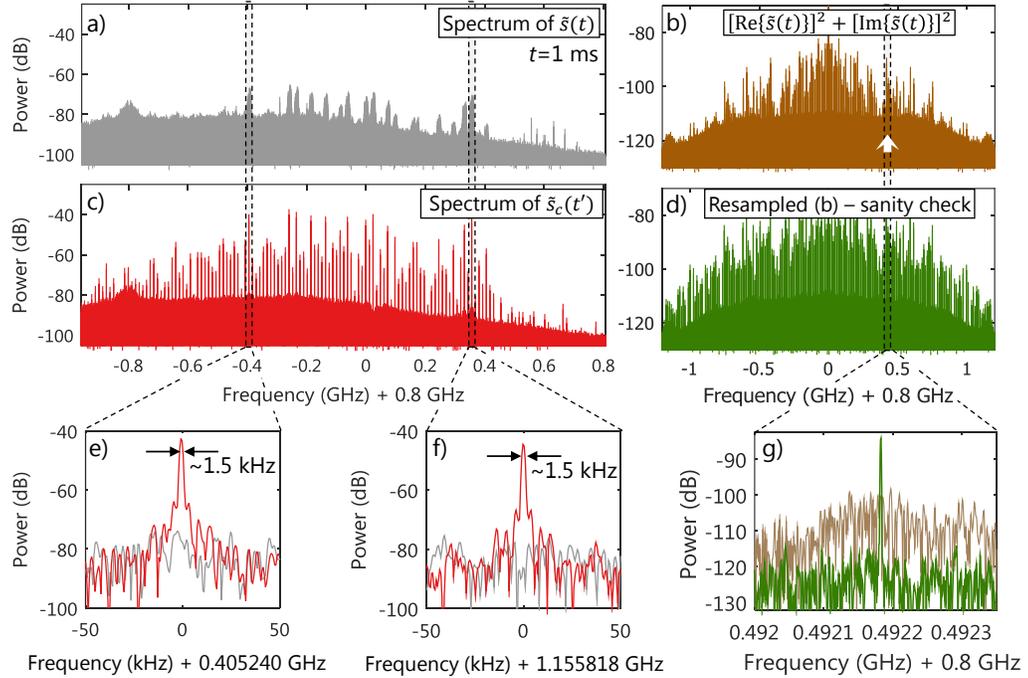

Fig. 7. (a) Dual-comb spectrum of two ICL combs acquired over 1 ms, where both are operated in the comb-regime. (b) the DDFG self-mixing spectrum with several $\Delta f_{\text{rep}}$-harmonics that can be used for the adaptive sampling procedure. (c) the rf spectrum after correction. (d) the resampled DDFG self-mixing spectrum. (e), (f) zoom of a rf beat note showing the effects of the CoCoA algorithm (g) zoom of a DDFG self-mixing harmonic showing the effects of the adaptive sampling.

## 5. Comparison to an analog phase-locked loop

It is interesting to note the similarities of the CoCoA algorithm to that of an optical phase-locked loop (OPLL), which actively stabilizes the frequency offset between the sources, thereby significantly reducing the difference in frequency offset. The effects of such a procedure implemented on a pair of long-wave infrared (LWIR) QCL combs [27] is shown in Fig. 8a and more details on the implementation of the phase-locking is given in ref. [22].

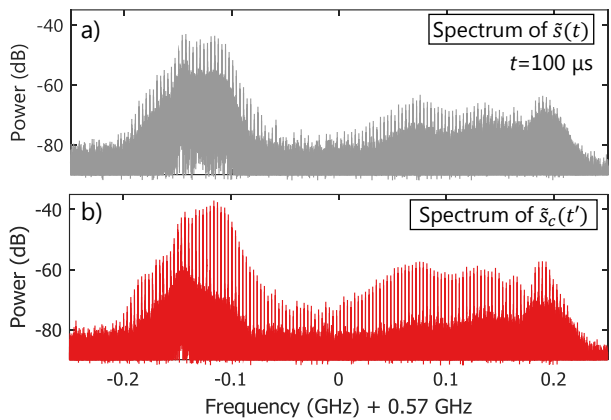

Fig. 8. (a) DCS rf spectrum using two LWIR QCL combs, phase-locked via active feedback control of the injection current. (b) DCS rf spectrum from the same sources using the CoCoA algorithm described in this work.

As seen in the figure, the OPLL produces resolvable beat notes especially close to the region where the phase-reference local oscillator is placed. However, there is no adaptive sampling of the interferograms and hence an increase in linewidth can be observed for higher rf frequency components.

The CoCoA algorithm clearly outperforms the analog OPLL across the spectrum and provides improved beat note signal-to-noise ratio without introducing any of the complex instrumentation that is associated with a high-bandwidth locking system. The benefits of phase-locking can thereby be obtained even for portable systems, where requirements on size, weight and power are imposed.

## 6. Summary

A novel and purely computational coherent averaging (CoCoA) algorithm for dual-comb spectroscopy is presented. The algorithm is compatible with modern real-time architectures and can be implemented without any additional electro-optical components, which simplifies the systems and lowers the cost. The CoCoA algorithm works directly on the acquired digitized photodetector signal and its efficacy is demonstrated on a spectrum acquired with two terahertz quantum cascade laser frequency combs [11,28] operated under conditions that introduce significant amounts of phase noise. Even though this serves as an illustrative example, the generality of the proposed solution is verified by application to other systems with varying degrees of phase noise. In principle, the procedure can be implemented in any DCS system, but the main benefits will be seen in on-chip semiconductor systems that exhibit intrinsically high degrees of phase fluctuations. If implemented, coherent averaging over long time-scales is feasible uncovering rf components otherwise hidden in noise.

**Appendix**

*Derivation of the digital difference frequency generation (DDFG) formula*

To describe how this procedure works mathematically, we will start with the resampling step first, which employs Digital Difference Frequency Generation (DDFG). Consider a radio-frequency $N$-component complex multiheterodyne signal in the form:

$$\tilde{s}(t) = \sum_{n=1}^{N} \cos\bigl(2\pi(\Delta f_0 + n\Delta f_{\text{rep}})t\bigr) + i \sum_{n=1}^{N} \sin\bigl(2\pi(\Delta f_0 + n\Delta f_{\text{rep}})t\bigr)$$
$$= \sum_{n=1}^{N} \exp\bigl(i2\pi(\Delta f_0 + n\Delta f_{\text{rep}})t\bigr), \qquad (8)$$

where $\Delta f_0$ is the rf offset frequency, and $\Delta f_{\text{rep}}$ is the rf repetition frequency. For simplicity, we assume equal amplitudes of the $N$ frequency components. Our goal is to retrieve $N$ harmonics of $\Delta f_{\text{rep}}$, while eliminating any frequency components containing $\Delta f_0$. This problem is almost equivalent to difference frequency generation (DFG) – a process well known from the domain of nonlinear optics [29]. In the optical domain, the second order nonlinear susceptibility $\chi^{(2)}$ of a medium is responsible for multiple frequency mixing phenomena, such as sum frequency generation (SFG), second harmonic generation (SHG), and difference frequency generation (DFG), all dependent upon the squared electric field. In principle, this scheme can be easily applied here to obtain the difference frequencies just by squaring the input signal. However, for a complex input we will just double and sum the signal's frequencies without creating any difference components. Using the well-known formula for the square of the sum of $N$ numbers, we can write:

$$\left(\sum_{n=1}^{N} a_n\right)^2 = \sum_{n=1}^{N} a_n^2 + 2 \sum_{m=1}^{N} \sum_{l=1}^{m-1} a_l a_m. \qquad (9)$$

If we insert Eq. (2) into the above equation, we obtain

$$[\tilde{s}(t)]^2 = \underbrace{\sum_{n=1}^{N} \exp\bigl(i2\pi(2\Delta f_0 + 2n\Delta f_{\text{rep}})t\bigr)}_{\text{SHG}} +$$
$$+ 2 \underbrace{\sum_{m=1}^{N} \sum_{l=1}^{m-1} \exp\bigl(i2\pi(2\Delta f_0 + (l+m)\Delta f_{\text{rep}})t\bigr)}_{\text{SFG}}. \qquad (10)$$

The result is complex and spectrally asymmetric but there is no DFG term. Only the aforementioned second harmonic, and sum frequency components appear in the equation.

Let us now analyze, what happens if we process the real and imaginary components separately. By squaring the real part of $\tilde{s}(t)$, we obtain the same components as given by $\chi^{(2)}$ in nonlinear optics:

$$(\text{Re}\{\tilde{s}(t)\})^2 = \underbrace{\frac{N}{2}}_{\text{DC}} + \underbrace{\frac{1}{2}\sum_{n=1}^{N}\cos\left(2\pi(2\Delta f_0 t + 2n\Delta f_{\text{rep}} t)\right)}_{\text{SHG}} +$$
$$+ \underbrace{\sum_{m=1}^{N}\sum_{l=1}^{m-1}\cos\left(2\pi(2\Delta f_0 t + (l+m)\Delta f_{\text{rep}} t)\right)}_{\text{SFG}} + \qquad (11)$$
$$+ \underbrace{\sum_{m=1}^{N}\sum_{l=1}^{m-1}\cos(2\pi(m-l)\Delta f_{\text{rep}} t)}_{\text{DFG}}.$$

In the above formula, we employed the product-to-sum formula for the cosine, namely $2\cos x \cos y = \cos(x+y) + \cos(x-y)$, where $x = 2\pi\Delta f_0$, and $y = 2\pi n\Delta f_{\text{rep}}$. Since the complex character of the signal is lost, the frequency spectrum becomes symmetric with a significant DC term. The latter in nonlinear optics is known as optical rectification, and here the same process takes place: by squaring a signal, we rectify it thus producing DC. More importantly, the DFG term appears now in the equation, yielding harmonics of $\Delta f_{\text{rep}}$. Unfortunately, they will often be overlaid by spectrally-broad low frequency SHG and SFG components containing $\Delta f_0$. To make matters worse, higher frequency SFG and SHG products can spectrally alias if they appear at frequencies greater than half of the sampling frequency, thereby spectrally overlaying the DFG products as well. Both effects will introduce a cross-talk between the DFG products used to retrieve correction signals, and the unwanted SHG and SFG by-products. Clearly, a nonlinear operation that yields DFG alone without any second harmonic or sum products including $\Delta f_0$ is of utmost importance.

If we now square the imaginary part of $\tilde{s}(t)$, the result is nearly the same, however the sign of the components containing $\Delta f_0$ flips:

$$(\text{Im}\{\tilde{s}(t)\})^2 = \underbrace{\frac{N}{2}}_{\text{DC}} - \underbrace{\frac{1}{2}\sum_{n=1}^{N}\cos(2\pi(2\Delta f_0 + 2n\Delta f_{\text{rep}})t)}_{\text{SHG}} +$$
$$- \underbrace{\sum_{m=1}^{N}\sum_{l=1}^{m-1}\cos(2\pi(2\Delta f_0 + (l+m)\Delta f_{\text{rep}})t)}_{\text{SFG}} + \qquad (12)$$
$$+ \underbrace{\sum_{m=1}^{N}\sum_{l=1}^{m-1}\cos(2\pi(m-l)\Delta f_{\text{rep}} t)}_{\text{DFG}}.$$

Note that the minus sign appears exactly where it is needed. If we combine the two squared parts of the complex signal, any influence on $\Delta f_0$ disappears, therefore we arrive at the final formula for multiple harmonics of the repetition rate of a multiheterodyne signal, referred here to as Digital Difference Frequency Generation (DDFG):

$$(\text{Re}\{\tilde{s}(t)\})^2 + (\text{Im}\{\tilde{s}(t)\})^2 = \underbrace{N}_{\text{DC}} + \underbrace{2\sum_{m=1}^{N}\sum_{l=1}^{m-1}\cos\bigl(2\pi(m-l)\Delta f_{\text{rep}}t\bigr)}_{\text{DFG}}. \quad (13)$$

This formula gives easy access to all harmonics of the repetition rate and allows for direct use of fast harmonic frequency trackers [30–32]. Alternatively, seeing that the fluctuations of the difference in repetition rates have a cumulative effect, we can pick a high-order repetition rate harmonic beat note, filter it using less computationally demanding low-order bandpass filters, retrieve its instantaneous frequency through IQ demodulation, and finally scale the estimated frequency by the harmonic number. In contrast to ultra-sharp narrowband filters used for retrieving the instantaneous repetition frequency of low-order beat notes close to DC, as only such are usable in signals contaminated by SHG and SFG products, the proposed high-order harmonic approach introduces less pronounced edge effects and ringing artefacts, which translates into more efficient correction of a noisy multiheterodyne signal.

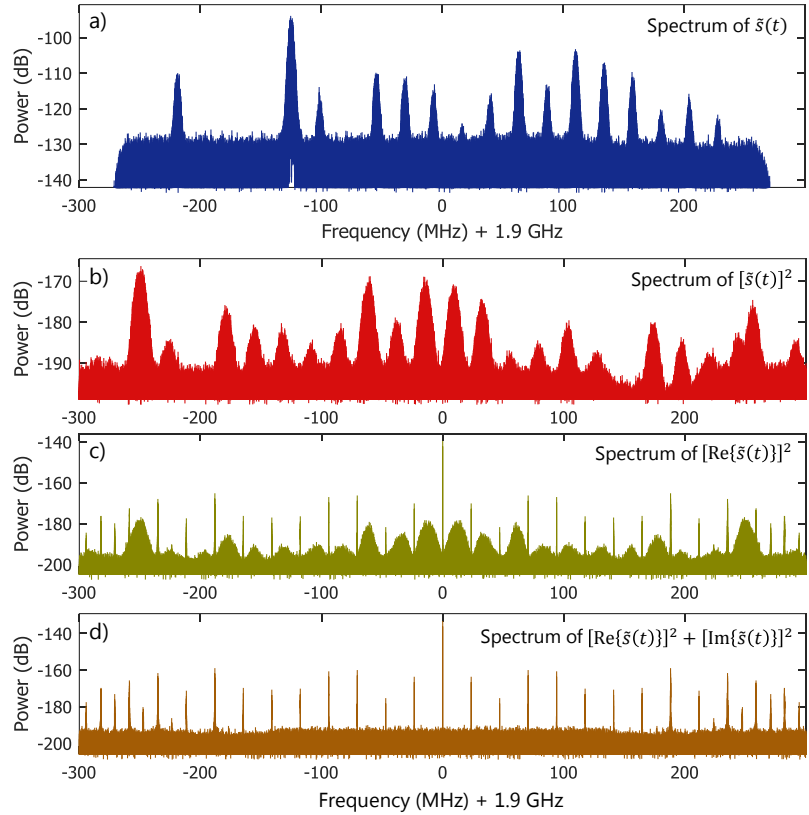

Fig. 9. Spectral comparison of the different nonlinear operations applied to the acquired complex multiheterodyne signal in (a). Panel (b) shows the result of squaring the complex signal, which yields SHG and SFG without any DFG products. Panel (c) shows the spectrum of the squared real part, which creates a DC component together with SHG, SFG, and DFG. Note the spectral symmetry, and overlap of the components. (d) Spurious-free DFG signal obtained by adding the independently squared real and complex part of the signal. Slight aliasing is also visible for the highest order repetition rate harmonics.

**Funding**

DARPA SCOUT program (W31P4Q161001), Thorlabs Inc. L. A. Sterczewski acknowledges support from the Kosciuszko Foundation Research Grant.

**Acknowledgments**

The authors would like to thank Prof. Jerome Faist at ETH and Dr. Mahmood Bagheri at JPL for providing the LWIR QCLs and the ICLs used in this work, and Prof. Qing Hu at MIT for providing the THz QCL combs. The authors would also like to thank Dr. David Burghoff for fruitful discussions and comments during the developments of the CoCoA algorithm.